\documentclass[conference]{IEEEtran}

\IEEEoverridecommandlockouts 

\usepackage[noadjust]{cite}
\usepackage{graphicx,color}
\usepackage[dvipsnames]{xcolor}
\usepackage{amsmath,amsbsy,amsfonts,amssymb}
\usepackage{mathrsfs}
\usepackage{mathtools}
\usepackage{flushend}
\usepackage{braket}   
\usepackage[nolist]{acronym}  
\usepackage[linesnumbered,ruled,vlined]{algorithm2e}  
\SetKwInput{Input}{Input}
\SetKwInput{Output}{Output}

\mathtoolsset{showonlyrefs}
\ifCLASSOPTIONcompsoc
\usepackage[caption=false,font=normalsize,labelfont=sf,textfont=sf]{subfig}
\else
\usepackage[caption=false,font=footnotesize]{subfig}
\fi
\usepackage{booktabs}

\usepackage{bm} 
\usepackage{tikz}
\usetikzlibrary{%
    positioning,
    shapes.geometric,
    arrows.meta,
    calc
}
\usepackage{pgfplots}
\usepackage[caption=false,font=footnotesize]{subfig}
\usepackage[normalem]{ulem}
\usepackage{placeins}

\newcommand{\vn}{\text{v}}
\newcommand{\cn}{\text{c}}

\newcommand{\mch}{l}
\newcommand{\mchv}{\bm{l}}

\newcommand{\app}{\lambda}
\newcommand{\appv}{\bm{\lambda}}

\newcommand{\mvc}{l_{\vn_j \rightarrow \cn_i}}
\newcommand{\mcv}{l_{\cn_i \rightarrow \vn_j}}

\newcommand{\mvcf}[2]{l_{\vn_{#1} \rightarrow \cn_{#2}}}
\newcommand{\mcvf}[2]{l_{\cn_{#1} \rightarrow \vn_{#2}}}
\newcommand{\itc}{\tau}

\newlength{\plotwidth}
\newlength{\plotheight}

\DeclareMathOperator*{\argmax}{arg\,max}

\definecolor{MAT1}{rgb}{0, 0.4470, 0.7410}
\definecolor{MAT2}{rgb}{0.8500, 0.3250, 0.0980}
\definecolor{MAT3}{rgb}{0.9290, 0.6940, 0.1250}
\definecolor{MAT4}{rgb}{0.4940, 0.1840, 0.5560}
\definecolor{MAT5}{rgb}{0.4660, 0.6740, 0.1880}
\definecolor{MAT6}{rgb}{0.3010, 0.7450, 0.9330}
\definecolor{MAT7}{rgb}{0.6350, 0.0780, 0.1840}

\makeatletter
\def\@texttop{\vskip 4pt}
\makeatother

\pgfplotsset{compat=1.18} 

\begin{document}

\begin{acronym}
    \acro{BP}{belief propagation}
    \acro{CSS}{Calderbank-Shor-Steane}
    \acro{LP}{linear programing}
    \acro{RS}{restart belief}
    \acro{OSD}{ordered statistics decoding}
    \acro{LLR}{log-likelihood ratio}
    \acro{RB}{restart belief}
    \acro{QEC}{quantum error correction}
    \acro{MWPM}{minimum-weight perfect matching}
    \acro{PCM}{parity-check matrix}
    \acro{LSD}{localized statistics decoding}
    \acro{LER}{logical error rate}

    \acro{BPGD}{BP with Guided Decimation}
\end{acronym}

\title{ Pathfinder-BP: A Low-Latency Recursive Decoder for Topological Codes}

\author{
\IEEEauthorblockN{Stefano Tinelli$^{\dagger\ddagger}$, Francisco L\'azaro$^\dagger$, Francisco Garc\'ia Herrero$^\ddagger$}
\IEEEauthorblockA{$^\dagger$Institute of Communications and Navigation, German Aerospace Center (\{Stefano.Tinelli,Francisco.LazaroBlasco\}@dlr.de) \\
$^\ddagger$Universidad Complutense de Madrid, Spain (francg18@ucm.es)}

\vspace{-1.1cm}

\thanks{The work of Stefano Tinelli and Francisco L\'azaro is part of the HESC lighthouse project within Munich Quantum Valley initiative and is supported by the Bavarian state government with funds from the Hightech Agenda Bavaria.
}
}
\date{}
\maketitle
\thispagestyle{empty}
\pagestyle{empty}

\begin{abstract}
We propose Pathfinder-BP, a recursive decoder for topological quantum codes that combines belief propagation with controlled bit-flipping. An initial BP pass identifies and permanently fixes high-confidence errors. Upon convergence failure, the decoder recursively explores alternative error patterns by tentatively flipping low-confidence variables. On surface codes, Pathfinder-BP achieves a noise threshold and logical error rate comparable to minimum-weight perfect matching (MWPM). A crucial advantage of Pathfinder-BP is its highly parallelizable structure, which yields a worst-case decoding latency quasi-linear in the code distance, making it a suitable candidate for low-latency hardware implementation. 
\end{abstract}

\begin{IEEEkeywords}
Quantum error correcting codes, message-passing decoder, QLDPC codes, bit-flip decoder. \end{IEEEkeywords}

\vspace{-2.5mm}
\section{Introduction}
Reliable quantum computation requires real-time decoding of quantum error-correcting codes. As system sizes scale, decoders must balance correction performance, computational complexity, and hardware implementability. This challenge is particularly acute for topological and LDPC codes, where syndrome extraction must be processed rapidly to prevent error accumulation.

State-of-the-art decoders like \ac{OSD} \cite{panteleev2021degenerate}, \ac{LSD} \cite{Hillmann_2025}, and Tesseract \cite{beni2025} achieve strong correction capabilities but rely on matrix inversion, clustering, or sequential search, rendering them impractical for real-time hardware deployment \cite{maurya2026}. Conversely, belief propagation (BP) and its variants \cite{muller2025improved,ostojic2010multitree,valentini2025,wang2025,yao2024belief} offer low complexity and high parallelism, but frequently fail to converge on topological codes due to short cycles in their Tanner graphs. 

To bridge this gap, we propose Pathfinder-BP, a decoder that augments BP accuracy with
a parallel search over speculative bit-flips. If an initial BP pass fails to
converge, high-confidence errors are permanently fixed, and the decoder
explores alternative error patterns by recursively flipping low-confidence
variables and re-solving the residual syndrome.
Pathfinder-BP belongs to the family of BP decoders with speculative
bit-flips~\cite{varnica2007augmented, ostojic2010multitree, rigby2019modified,
wang2025, valentini2025}, from which it inherits the parallel branch
structure, the nesting of BP calls, and the residual decoding problem.
Its novelty lies in how the history of the initial BP stage steers the search.
First, the sign-change history of the posterior beliefs, used
in~\cite{wang2025} to select tentative flips, is used here for decimation:
variables whose belief crossed the decision boundary exactly once and then
stabilized are fixed before branching, shrinking the problem shared by all
branches.
Second, as in~\cite{ostojic2010multitree}, tentative errors are drawn from the
neighbourhood of unreliable checks, but reliability is measured by the number
of BP iterations during which a check remains unsatisfied.

Although Pathfinder-BP applies to arbitrary CSS codes, we focus here on surface codes under code-capacity noise, where it performs on par with MWPM while guaranteeing, unlike relay BP, a bounded latency for every syndrome, whose worst case grows quasi-linearly in the code distance.

\vspace{-1mm}
\section{Preliminaries}

\vspace{-0.5mm}
\subsection{Stabilizer Codes}
An $[[n,k,d]]$ stabilizer code encodes $k$ logical qubits into $n$ physical qubits and is specified by $m=n-k$ independent, commuting $n$-qubit Pauli operators $\bm{S}_1,\dots,\bm{S}_m$, the generators of the stabilizer group $\boldsymbol{\mathcal{S}}$  and has a code distance $d$. The code space $\boldsymbol{\mathcal{C}}$ is the set of states $\ket{\phi}$ satisfying $\bm{S}_i\ket{\phi}=\ket{\phi}$ for all $i$. If a Pauli error $\bm{E}$ acts on $\ket{\phi}\in\boldsymbol{\mathcal{C}}$, measuring $\bm{S}_i$ leaves $\bm{E}\ket{\phi}$ unaltered and yields $s_i=0$ if $\bm{E}$ commutes with $\bm{S}_i$, and $s_i=1$ otherwise. The length-$m$ binary vector $\bm{s}=(s_1,\dots,s_m)$ is the syndrome.
 
An $n$-qubit Pauli operator is represented by a binary vector $(\bm{x}|\bm{z})$ of length $2n$, where $x_i=1$ if the $i$-th factor is $\bm{X}$ or $\bm{Y}$, and $z_i=1$ if it is $\bm{Z}$ or $\bm{Y}$. Stacking the representations of the generators yields the $m\times 2n$ parity-check matrix $\bm{H}=(\bm{A}|\bm{B})$. In this work we consider \ac{CSS} codes \cite{Calderbank_1996}, whose generators contain either only $\bm{X}$ or only $\bm{Z}$ factors, so that $\bm{H} = \begin{bmatrix} \bm{H}_X & \bm{0} \\ \bm{0} & \bm{H}_Z \end{bmatrix}$. Note that $\bm{H}$ is not unique, and it may contain $m^\prime>m$ rows, $m^\prime-m$ of them being linearly dependent.

\subsection{Error Model}
\label{sec:err_model}
We consider code capacity noise, i.e., syndrome extraction is assumed to be perfect. The error is described by a binary vector $\bm{e}=(\bm{e}_X|\bm{e}_Z)$ of length $2n$ \cite{MacKay_2004}, and the syndrome is $\bm{s}^\intercal=\bm{H}\bm{M}\bm{e}^\intercal$, where $\bm{M} = \begin{bmatrix} \bm{0}_n & \bm{I}_{n} \\ \bm{I}_n & \bm{0}_n \end{bmatrix}$ swaps the two halves of $\bm{e}$, $\bm{M}\bm{e}^\intercal=(\bm{e}_Z|\bm{e}_X)^\intercal$. The physical qubits are subject to independent $\bm{X}$ and $\bm{Z}$ errors with equal probability,
\begin{equation}
    \label{eq:prob}
    \Pr\{e_i=1\}=\epsilon, \quad \forall i \in \{1,\dots,2n\}.
\end{equation}
For \ac{CSS} codes the syndrome splits as $\bm{s}=(\bm{s}_X|\bm{s}_Z)$, with $\bm{s}_X^\intercal=\bm{H}_X\bm{e}_Z^\intercal$ and $\bm{s}_Z^\intercal=\bm{H}_Z\bm{e}_X^\intercal$, so that $\bm{X}$ and $\bm{Z}$ errors can be decoded independently. In the following we consider $\bm{Z}$ errors only, the decoding of $\bm{X}$ errors being identical, and write $\bm{e}=\bm{e}_Z$ and $\bm{H}=\bm{H}_X$.

\section{The Decoding Problem}
Given the observed syndrome $\bm{s}$ and a probabilistic model for $\bm{e}$, the task of the decoder is finding the most likely error vector $\hat{\bm{e}} = \argmax_{ \bm{e} \bm{H}^\intercal  = \bm{s} } p(\bm{e}|\bm{s})$
that is compatible with observed syndrome  $\bm{s}$.
Under the assumption that the qubits are affected by i.i.d. bit and phase flips (X and Z errors), this is equivalent to finding the vector $\hat{\bm{e}}$ of lowest possible Hamming weight that yields that syndrome:
\begin{equation}
    \label{eq:decProblminW}
        \hat{\bm{e}} = \arg \min_{\bm{e}   \bm{H}^\intercal  = \bm{s} } \left( \sum_i e_i \right) 
\end{equation}

This decoding problem is, in general, NP-hard, making exact solutions computationally intractable for large codes. Notable exceptions are topological codes, such as the surface and toric codes, whose structure allows for efficient exact decoding. 
For these codes, \ac{MWPM} returns a minimum-weight error consistent
with the syndrome in polynomial time \cite{Higgott_2025}, and hence
solves Eq.~\eqref{eq:decProblminW} exactly\footnote{We remark that
Eq.~\eqref{eq:decProblminW} is actually not the optimal decoding criterion. The optimal decoder selects the most probable
equivalence class of errors (i.e., it sums the probabilities of all
errors differing by a stabilizer) rather than the most probable
individual error vector \cite{iyer2015hardness}.}. 

\subsection{Belief Propagation}
Under the assumption of independent noise, it is possible to approximate $ \hat{\bm{e}}$ by $\tilde{\bm{e}}$, where $\tilde{{e}}_i = \argmax_{e_i} p(e_i|\bm{s}).$
A good approximation of $\tilde{\bm{e}}$ can be obtained using \ac{BP} decoding, which is best understood relying on a graphical representation of the parity check matrix $\bm{H}$ (the Tanner graph), where variable and check nodes exchange messages along the edges.
Denoting by $\mvc^{(\itc)}$ the message sent from variable node $\vn_j$ to check node $\cn_i$ at the $\itc$-th iteration, we have
\begin{equation}
    \label{VNbin}
        \mvc^{(\itc)} =\mch_j +\sum_{i'|\cn_{i'}\in \mathcal{N}(\vn_j)\setminus \cn_i} \mcvf{i'}{j}^{(\itc)}
\end{equation}
where $\mathcal{N}(\vn_j)$ denotes the set of neighbours of $\vn_j$ and $\mch_j = \log\left(\frac{1-\epsilon}{\epsilon}\right)$ is the channel log-likelihood ratio of $\vn_j$. The check-to-variable messages are computed as
\begin{equation}
    \label{CNbin}
        \mcv^{(\itc)} = (-1)^{s_i} \, 2 \, \tanh^{-1} \mkern-5mu\left( \prod_{j'|\vn_{j'}\in  \mathcal{N}(\cn_i) \setminus v_j} \mkern-15mu \tanh\left( \frac{1}{2} \mvcf{j'}{i}^{(\itc-1)} \right)\right).
\end{equation}
The posterior LLRs at the end of the $\itc$-th iteration are
\begin{equation}
    \label{LLRMarBin}
        \app_j^{(\itc)} = \mch_j +\sum_{i|\cn_i\in \mathcal{N}(\vn_j)} \mcv^{(\itc)}.
\end{equation}
Based on the posterior \acp{LLR}, the hard-decision estimates at iteration $\itc$ are obtained as
\begin{equation}
    \begin{array}{cc}
        \tilde{e}_j^{(\itc)} = 0 & \mbox{if  } \app_j^{(\itc)}\ge0 \\
        \tilde{e}_j^{(\itc)} = 1 & \mbox{if  } \app_j^{(\itc)}<0 
    \end{array}
\end{equation}
with associated syndrome $ \tilde{ \bm{s} }= \tilde{\bm{e}} \bm{H}^\intercal $.
BP runs until a predefined maximum number of iterations is reached or until it converges to an error pattern compatible with the observed syndrome, i.e., until $\tilde{\bm{s}}=\bm{s}$.
\ac{BP} is known to be optimal when the Tanner graph is cycle-free. However, the Tanner graphs of many stabilizer codes contain short cycles and, in addition, degenerate errors split the belief among equivalent solutions. Both effects severely impair the performance of BP \cite{poulin2008}. The most common failure occurs when \ac{BP} is unable to find an error pattern compatible with the observed syndrome.

\vspace{-2mm}
\subsection{Prior Work}
\label{sec:pri_work}
In the following we present different techniques that have been proposed to overcome  this limitations ob BP decoding for QEC codes.

In the context of quantum error correction codes, \ac{OSD} was first proposed in \cite{panteleev2021degenerate} as a postprocessing stage to BP that is guaranteed to yield an error compatible with the observed syndrome.
\ac{OSD} reorders the bit indices according to the reliability of the BP posterior LLRs $\bm{\appv}$ and partitions the reordered parity-check matrix into a basis $\mathcal{A}$ of $r=\mathrm{rank}(\bm{H})$ linearly independent columns and a remainder set $\mathcal{T}$. In order-$\ell$ OSD, all error patterns on $\mathcal{T}$ with Hamming weight $\le \ell$ are enumerated, and for each of them the corresponding basis error is obtained by solving a linear system involving $\bm{H}_{[\mathcal{A}]}$, yielding an error that satisfies the syndrome. The minimum-weight candidate is then permuted back to the original ordering. The dominant cost is the inversion of $\bm{H}_{[\mathcal{A}]}$, which scales cubically with the code length and complicates hardware implementation.

\ac{BPGD} \cite{yao2024belief,gong2024} executes successive BP instances of $T_{\text{BP}}$ iterations each, where every instance adopts the output \acp{LLR} of the preceding one as input \acp{LLR}. When an instance fails to converge, the variable with largest absolute \ac{LLR} is decimated, i.e., its prior is set to $\pm\infty$ according to the sign of $\app_j$. The algorithm terminates when a BP instance converges to a valid solution or when all qubits have been decimated, whichever occurs first.

Relay BP \cite{muller2025improved} also uses a sequence of \ac{BP} instances to enhance convergence. After an initial \ac{BP} stage, if it fails, each subsequent stage $h$ computes its input channel probabilities as a damped combination of the original and the previous output \ac{LLR}s, $p_{\text{in}}^h = (1 - \gamma) p_{\text{in}}^0 + \gamma p_{\text{out}}^{(h - 1)}$, with the damping parameter $\gamma$ sampled uniformly from a code-specific distribution.
The algorithm runs until $\text{n}_{\text{cand}}$ candidate solutions have been collected or a maximum number of \ac{BP} stages has been executed, and finally selects the minimum-weight candidate.

Several works in the literature augment BP by introducing speculative bit-flips \cite{ostojic2010multitree, wang2025, valentini2025}. A first instance of BP is run and, if it fails to converge, a list of $\eta$ tentative errors is built, i.e., variable nodes that are likely to be in error. In a second stage, $\eta$ decoding branches are launched (possibly in parallel). Here, the $i$-th branch flips the $i$-th tentative error (e.g., setting its LLR to $-\infty$) and attempts BP decoding again. Finally, the most likely error among the solutions of the different branches is chosen. These works differ, among other aspects, in the way the tentative errors are identified. The earliest example is \cite{ostojic2010multitree}, which considers classical LDPC codes: a reliability metric is defined for the check nodes, and tentative errors are identified as the least reliable neighbours of the least reliable check nodes. In \cite{wang2025} the tentative errors are the variable nodes whose beliefs change sign most frequently during the first BP stage. In restart-belief \cite{valentini2025} they are the $\eta$ variable nodes with lowest posterior LLRs; if the second BP stage fails to converge, a further tentative error is flipped (again the variable node with lowest posterior LLR), and BP is attempted again until convergence or until a maximum number of nested BP calls is reached.

\vspace{-2mm}
\section{Pathfinder-BP Decoder}\label{sec:Pathfinder-BP}
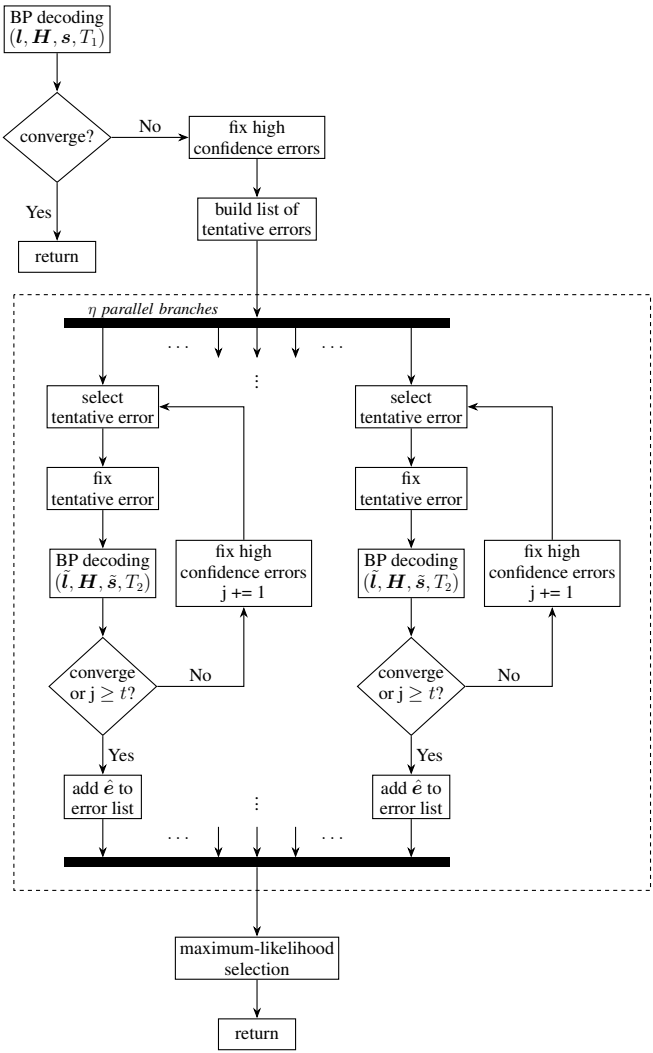
\begin{figure}
    \centering
    \scalebox{0.51}{
    \begin{tikzpicture}[
    font=\fontsize{13.5}{15}\selectfont,
    node distance = 1.0cm and 1.5cm,
    base/.style = {draw, minimum width=2cm, minimum height=0.8cm, align=center, thick, fill=white},
    process/.style = {base},
    decision/.style = {diamond, draw, minimum width=2.8cm, minimum height=1.4cm, align=center, thick, inner sep=0pt,  fill=white},
    arrow/.style = {-{Stealth[scale=1.2]}, thick}
]

\node [process] (start) {BP decoding\\$(\mchv, \bm{H}, \bm{s}, T_1)$};
\node [decision, below=of start] (dec1) {converge?};
\node [process, below=of dec1, yshift=-0.5cm] (return_main) {return};
\node [process, right=of dec1, xshift=0.5cm] (blockA) {fix high \\confidence errors};
\node [process, below=of blockA] (blockB) {build list of \\ tentative errors };

\node[fill, black, minimum width=10cm, minimum height=0.1cm, below=of blockB, yshift=-1cm] (fork) {};

\node [process, below=of fork, xshift=-4.0cm, yshift=-0.5cm] (c1_1) {select \\ tentative error};
\node [process, below=of c1_1] (c2_1) {fix \\ tentative error};
\node [process, below=of c2_1] (c3_1) {BP decoding\\$(\Tilde{\mchv}, \bm{H}, \Tilde{\bm{s}}, T_2)$};
\node [decision, below=of c3_1] (dec_c1_1) {converge \\ or j  $\ge t$?};
\node [process, below=of dec_c1_1] (add1_1) {add $\hat{\bm{e}}$  to  \\error list};
\node [right=of dec_c1_1] (add1_1_r) {};
\node [process, right=of c3_1, , xshift=-1cm] (c4_1) {fix high \\confidence errors \\ j += 1};

\node [process, below=of fork, xshift=4.0cm, yshift=-0.5cm] (c1_2) {select \\ tentative error};
\node [process, below=of c1_2] (c2_2) {fix \\ tentative error};
\node [process, below=of c2_2] (c3_2) {BP decoding\\$(\Tilde{\mchv}, \bm{H}, \Tilde{\bm{s}}, T_2)$};
\node [decision, below=of c3_2] (dec_c1_2) {converge \\ or j $\ge t$?};
\node [process, below=of dec_c1_2 ] (add1_2) {add $\hat{\bm{e}}$  to  \\error list};
\node [process, right=of c3_2, , xshift=-1cm] (c4_2) {fix high \\confidence errors\\ j += 1};

\node[fill, black, minimum width=10cm, minimum height=0.1cm, below=of add1_1, yshift=-0cm, xshift=4.0cm] (join) {};

\begin{scope}
    \draw[dashed, thick] ($(fork.north west)+(-1.3, 0.6)$) rectangle ($(join.south east)+(5.2, -0.6)$);
    \node[anchor=north west, font=\itshape] at ($(fork.north west)+(0.5, 0.5)$) {\large{$\eta$ parallel branches }};
\end{scope}

\node [process, below=of join, yshift=-0.8cm] (blockD) {maximum-likelihood \\selection};
\node [process, below=of blockD] (blockE) {return};

\draw [arrow] (start) -- (dec1);
\draw [arrow] (dec1) -- node[left] {Yes} (return_main);
\draw [arrow] (dec1) -- node[above] {No} (blockA);
\draw [arrow] (blockA) -- (blockB);
\draw [arrow] (blockB) -- (fork);

\draw [arrow] (fork.south -| c1_1) -- (c1_1);
\draw [arrow] (c1_1) -- (c2_1); 
\draw [arrow] (c2_1) -- (c3_1);
\draw [arrow] (c3_1) -- (dec_c1_1);
\draw [arrow] (dec_c1_1) -- node[right] {Yes} (add1_1); 
\draw [arrow] (add1_1) -- (add1_1 |- join.north);

\draw [arrow] (dec_c1_1.east) -| node[pos=0.25, above] {No} (c4_1.south);
\draw [arrow] (c4_1.north) |- (c1_1.east);

\draw [arrow] (fork.south -| c1_2) -- (c1_2);
\draw [arrow] (c1_2) -- (c2_2); 
\draw [arrow] (c2_2) -- (c3_2);
\draw [arrow] (c3_2) -- (dec_c1_2);
\draw [arrow] (dec_c1_2) -- node[right] {Yes} (add1_2); 
\draw [arrow] (add1_2) -- (add1_2 |- join.north);

\draw [arrow] (dec_c1_2.east) -| node[pos=0.25, above] {No} (c4_2.south);
\draw [arrow] (c4_2.north) |- (c1_2.east);

\draw [arrow] (join) -- (blockD);
\draw [arrow] (blockD) -- (blockE);

\draw [arrow] ([xshift=-1.0cm]fork.south) -- ++(0,-0.8);
\draw [arrow] ([xshift=0cm]fork.south) -- ++(0,-0.8);
\draw [arrow] ([xshift=1.0cm]fork.south) -- ++(0,-0.8);

\node at ($(fork.south)+(-2.0, -0.5)$) {\Large $\dots$};
\node at ($(fork.south)+(+2.0, -0.5)$) {\Large $\dots$};

\node at ($(fork.south) + (0,-1.3)$) {\Large $\vdots$}; 

\draw [arrow] ([xshift=-1.0cm]join.north) ++(0,0.8) -- ([xshift=-1.0cm]join.north);
\draw [arrow] ([xshift=0cm]join.north) ++(0,0.8) -- ([xshift=0cm]join.north);
\draw [arrow] ([xshift=1.0cm]join.north) ++(0,0.8) -- ([xshift=1.0cm]join.north);

\node at ($(join.north)+(-2.0, +0.5)$) {\Large $\dots$};
\node at ($(join.north)+(+2.0, +0.5)$) {\Large $\dots$};

\node at ($(join.north) + (0,+1.5)$) {\Large $\vdots$}; 

\end{tikzpicture}
    }
    \caption{Schematic overview of the decoder architecture. The flowchart illustrates the sequential processing steps of the Pathfinder-BP decoder.}
    \label{fig:flow}
\end{figure}
\vspace{-1mm}
\subsection{Overview}\label{subsec:overview}
Pathfinder-BP is a decoding algorithm belonging to the family of \ac{BP} decoders with speculative bit-flips. Its overall structure is depicted in Fig.~\ref{fig:flow} and summarized in Algorithm~\ref{myalg}. As in \cite{ostojic2010multitree, wang2025, valentini2025}, decoding starts with an initial \ac{BP} stage, which runs for at most $T_1$ iterations. If this stage converges, its output is returned and no further processing takes place.

If the initial stage fails to converge, the decoder exploits the output of the initial BP decoding in two distinct ways:
\begin{enumerate}
    \item A first heuristic identifies a set $\mathcal{I}$ of \emph{high-confidence estimated errors}, i.e., variable nodes that are in error with high confidence. These nodes are \emph{fixed}, i.e., they are removed from the inference problem and their contribution is subtracted from the syndrome. This step is reminiscent of guided decimation \cite{yao2024belief}, but it differs both in the heuristic used and in the way the errors are fixed.
    \item A second heuristic identifies a list $\mathcal{L}$ of $\eta$ \emph{tentative errors}, i.e., variable nodes whose posterior beliefs remain uncertain after the initial \ac{BP} stage.
\end{enumerate}

The decoder then launches $\eta$ decoding branches, that can be executed in parallel. The $i$-th branch fixes the $i$-th tentative error $\mathcal{L}[i]$ on top of the already fixed high-confidence errors, and runs a \ac{BP} instance for $T_2$ iterations on the resulting residual decoding problem. If it converges, the branch reconstructs the corresponding candidate error pattern and terminates. Otherwise, the branch selects high-confidence errors and one further tentative error, fixes them, and invokes \ac{BP} again. Each branch is thus a nested sequence of \ac{BP} calls that progressively fixes error positions, terminating upon convergence or after a maximum recursion depth $t$ has been reached.

Finally, the candidate error patterns collected by all branches are gathered in a list $\mathcal{E}$, and the most likely one is returned. The two parameters $\eta$ and $t$ control, respectively, the width and the depth of the search, and  determine the trade-off between error-correction performance and complexity (see Section~\ref{sec:num_res}).

\vspace{-2mm}
\subsection{Initial BP stage}
\label{subsec:softinfo}
Besides the hard-decision estimate $\hat{\bm{e}}$ and the posterior \ac{LLR} vector $\appv$, defined in \eqref{LLRMarBin}, the initial \ac{BP} also keeps track of two quantities. The first one is the number of iterations during which each parity check remains unsatisfied. Specifically, we define the vector $\bm{\Theta}$ of length $m$, whose $i$-th entry corresponds to
\begin{equation}
\label{eq:theta}
\Theta_i = \left| \left\{ \tau \in \{1, \dots, T_1\} : \tilde{s}_i^{(\tau)} \neq s_i \right\} \right|.
\end{equation}
Thus, $\Theta_i$ counts the number of iterations in which the $i$-th parity check is violated during the initial $T_1$ iterations of \ac{BP}. Tracking $\bm{\Theta}$ requires only a small counter per check node and is therefore inexpensive to implement in hardware.

The quantity $\bm{\Theta}$ provides information about the reliability of the \ac{BP} decoding process, and allows us to identify problematic neighborhoods in the Tanner graph. In particular, a large value of $\Theta_i$ indicates that the corresponding check node either needed many BP iterations to converge or exhibited  oscillatory behaviour.

The second quantity counts the number of times that the posterior \ac{LLR} of $\vn_j$ changed sign over the $T_1$ iterations,
\begin{equation}
    \label{eq:phi}
    \Phi_j = \left| \left\{ \tau \in \{1,\dots,T_1\} : \mathrm{sign}\!\left(\app_j^{(\tau)}\right) \neq \mathrm{sign}\!\left(\app_j^{(\tau-1)}\right) \right\} \right|,
\end{equation}
where $\app_j^{(0)} = \mch_j$ denotes the channel LLR of $v_j$.
A large value of $\Phi_j$ indicates that $v_j$ is involved in short cycles
that keep oscillating its belief, the primary failure mechanism impairing
\ac{BP} on topological codes~\cite{poulin2008}. Tracking $\bm{\Phi}$ requires
one sign bit and a small counter per variable node, and is therefore
essentially free in hardware.

\vspace{-1mm}
\subsection{Identifying high-confidence errors}\label{subsec:highconf}
The set $\mathcal{I}$ of high-confidence errors collects the variable nodes that changed sign exactly once and whose final posterior \ac{LLR} is sufficiently negative, $\mathcal{I} = \left\{ j : \Phi_j = 1 \ \wedge \ \app_j \leq \alpha \right\}$, with $\alpha \leq 0$ a threshold parameter that needs to be optimized for each code. In our experiments with surface codes, setting $\alpha = -d$, with $d$ the distance of the code, yields good results across all the considered distances.
The condition $\Phi_j = 1 $ selects variable nodes whose belief crossed the decision boundary once and then stabilized, distinguishing genuine errors from oscillating variable nodes $\Phi_j \geq 2$ or from stable correct variable nodes $\Phi_j = 0$.

\subsection{Fixing errors}
\label{subsec:fixing}
In order to fix errors,  we introduce a \emph{forced-error vector} $\bar{\bm{e}}$, initialized to the all-zero vector, and a copy $\bar{\mchv}$ of the channel \ac{LLR} vector, initialized to $\bar{\mchv} = \mchv$. Fixing an error in position $a$ implies setting $\bar{e}_a = 1$, $\bar{\mch}_a = +\infty$
and computing a new \emph{target syndrome} $\bar{\bm{s}} = \bm{s} \oplus \bar{\bm{e}} \bm{H}^\intercal$.
The subsequent \ac{BP} call is run with $\bar{\mchv}$ as input \acp{LLR} and $\bar{\bm{s}}$ as observed syndrome, i.e., it solves the residual decoding problem for $\tilde{\bm{e}} = \bm{e} \oplus \bar{\bm{e}}$.
Whenever a \ac{BP} call converges to a residual estimate $\hat{\bm{e}}$ satisfying $\hat{\bm{e}}\bm{H}^\intercal = \bar{\bm{s}}$, the candidate for the original problem is recovered as $\hat{\bm{e}} \leftarrow \hat{\bm{e}} \oplus \bar{\bm{e}}$, which matches the observed syndrome $\bm{s}$.

\subsection{Building the list of tentative errors}\label{subsec:list}

After the initial BP stage, the list $\mathcal{L}$ is built as follows. The check nodes are sorted by decreasing $\Theta_i$, yielding a permutation $\bm{\pi}$ with $\Theta_{\pi(1)} \geq \cdots \geq \Theta_{\pi(m)}$. The check nodes are then visited in this order and, for each of them, its two neighbouring variable nodes with the lowest posterior \ac{LLR} magnitude are appended to $\mathcal{L}$, until $\eta$ (different) entries (variable nodes) have been collected. The tentative errors are thus the least reliable variable nodes in the neighbourhood of the checks that remained unsatisfied most often during the initial BP stage. At subsequent recursion levels, $\bm{\Theta}$ is recomputed, and the next tentative error is selected uniformly at random among the two least reliable neighbours of the check node with the largest $\Theta_i$.

{\setlength{\textfloatsep}{4pt}
\begin{algorithm}[t]
\small
\caption{ $[\hat{\bm e} ] \gets$ \texttt{Pathfinder-BP} ($\mchv, \bm{H}, \bm{s}, t, \eta)$ } %
\label{myalg}
	\Input{\\
		$\mchv$ — initial LLRs associated to the variable nodes \;
		$\bm{H}$ — parity check matrix  of the code\;
		$\bm{s}$ — observed syndrome\;
		$t$ — maximum recursion depth\;
		$\eta$ —  maximum list size\;
	}
	\Output{
		$ \hat{\bm e}$ — error estimate\;
	}
	\BlankLine

	 $[\hat{\bm e}, \appv , \bm{\Theta}, \bm{\Phi}] \gets \texttt{BP}(\mchv, \bm{H}, \bm{s})$\;
	 $\hat{\bm{s}} = \hat{\bm e} \bm{H}^\intercal $ \;

	\If{ $\hat{\bm{s}} =\bm{s}$}
	    {\Return{$\hat{\bm e}$ }}

	$\mathcal{I} \gets$ \texttt{SelectHighConfidErrors} $(\appv, \bm{\Phi})$ \;

	$\bar{\bm{e}}=\bm{0}$, \quad
	$\bar{\mchv} = \mchv$, \quad 
	$\mathcal{E} \gets \{\}$ \;

	$ [ \bar{\bm{e}}, \bar{\mchv} ] =$ \texttt{FixErrors}  $(\mathcal{I}, \emptyset, \bar{\bm{e}}, \bar{\mchv})$ \;

	$ [ \mathcal{L} ] =$ \texttt{GetListTentatErrors}  $( \bar{\bm{e}}, \appv, \bm{\Theta})$ \;

	\For{$i\gets 1$ \KwTo $\eta$ } 
	{
	    \textcolor{gray}{$/ /$ this loop is parallelizable }\textcolor{white}{\;}
		$\tilde{\mchv} = \bar{\mchv}$, \quad
		$\tilde{\bm{e}} = \bar{\bm{e}}$ \;

			\For{$j\gets 1$ \KwTo $t$}
			{
			\If{$j=1$}
			{
			$\breve{e} = \mathcal{L}[i]$ \;
			}
			\Else{
			$\breve{e} =$ \texttt{SelectTentatError} $(\tilde{\bm{e}} , \appv , \bm{\Theta}) $\;
			}
		    $ [ \tilde{\bm{e}}, \tilde{\mchv} ] =$ \texttt{FixErrors}  $(\mathcal{I}, \breve{e}, \tilde{\bm{e}}, \tilde{\mchv})$ \;
			$\tilde{\bm{s}} = \bm{s} \oplus \tilde{\bm{e}} \bm{H}^\intercal $\;
			$[\hat{\bm e}, \appv , \bm{\Theta}, \bm{\Phi}] \gets \texttt{BP}(\tilde{\mchv}, \bm{H}, \tilde{\bm{s}})$ \;
			$\hat{\bm{s}} = \hat{\bm e} \bm{H}^\intercal $ \;
			\If{$\hat{\bm{s}} =  \tilde{\bm{s}}$}
			{
			$\hat{\bm{e}} \gets \hat{\bm{e}} \oplus \tilde{\bm{e}}$\;
			$\mathcal{E} \gets \mathcal{E} \cup \{\hat{\bm{e}}\}$\;
			\textbf{break}\;
			}
			$\mathcal{I} \gets$ \texttt{SelectHighConfidErrors} $(\appv, \bm{\Phi})$ \;
			}
	}
    \textcolor{gray}{$/ /$ Among the solutions in the list, select the most likely one }\textcolor{white}{\;}
	$\hat{\bm{e}} \gets \arg\min_{\bm{e} \in \mathcal{E}} \sum_j e_j$\;
	{\Return{$\hat{\bm e}$ }}
\vspace{-1mm} \end{algorithm}

\subsection{Complexity and Latency}
Pathfinder-BP executes at most $1+\eta t$ BP instances. A flooding iteration sends one message per edge in each direction, i.e.\ exactly $2|\bm{H}|$ messages, where $|\bm{H}|$ denotes the number of nonzero entries of $\bm{H}$, with $|\bm{H}|\le 2n$ for the planar surface code and $|\bm{H}| = O(n)$ for any LDPC family with bounded degrees.

We derived latency in BP iterations on the worst-case scenario, assuming fully parallel BP cores.
Each recursion step additionally requires two global operations over all nodes (testing convergence and finding the check with largest $\Theta_i$), and the final selection one more, each costing $O(\log n)$ logic levels in hardware. The number of clock cycles is thus  $O\!\left(T_1 + tT_2+\log n\right)$.

For surface codes $n \sim d^2$. Setting $\eta = t = 2d$ (Sec.~\ref{sec:num_res}), the worst-case latency is $T_1 + t\,T_2 = 50+30d$ BP iterations, at the cost of $O(d^3)$ processing elements,  hence the worst-case number of operations is $O(d^4T_2)$. Both expressions assume $T_1 = 50$ and $T_2 = 15$ which is sufficed for all distances considered here.
\\
For comparison, the blossom algorithm runs sequentially with worst-case time $O(N^3)$ in the number $N \le m = O(d^2)$ of detection events, i.e.\ $O(d^6)$. Parallel implementations reduce this substantially: Micro Blossom~\cite{microblossom} employs $O(d^3)$ processing units to lower the worst-case latency from $O(d^{12})$ to $O(d^9)$. These exponents refer to a circuit-level decoding graph with $O(d^3)$ nodes and are therefore not directly comparable to our setting.
The relevant distinction is not the average latency, at which matching decoders excel, but its variability: the number of alternating-tree operations performed by blossom depends on the observed syndrome, whereas Pathfinder-BP terminates after at most $T_1+tT_2$ BP iterations for \emph{every} syndrome. A guaranteed worst-case bound is what a real-time decoder requires, since sporadic slow decoding events accumulate into a measurement backlog~\cite{fusionblossom}.

\section{Numerical Results}
\label{sec:num_res}

\begin{figure*}[t]
    \centering
    \begin{minipage}[t]{0.31\textwidth}
        \centering
        \subfloat[Pathfinder-BP with $\eta=t=2d$, $d\in\{3,5,\dots,15\}$, with the \ac{MWPM} threshold for reference.\label{fig:threshold}]{%
            \resizebox{\linewidth}{!}{%
                \begin{tikzpicture}
\pgfplotsset{
  set layers,
  mark layer=axis tick labels
}

\begin{loglogaxis}[
    width=\plotwidth,
    xmin=0.01,
    xmax=0.15,
    xlabel={$\epsilon$},
    ymin=1e-6,
    ymax=1,
    grid=both,   
    scaled x ticks=false,
    scaled y ticks=false,
    xtick={0.01,0.03,0.05,0.10,0.15},
    xticklabel={%
      \pgfkeys{/pgf/fpu=true}%
      \pgfmathparse{exp(\tick)}%
      \pgfmathprintnumber[fixed, precision=2, fixed zerofill]{\pgfmathresult}%
    },
    yticklabel={%
      \pgfmathparse{int(round(\tick/ln(10)))}%
      \ifnum\pgfmathresult=0 $1$\else $10^{\pgfmathresult}$\fi
    },
    ylabel={Logical error rate},
    legend columns=1,
    legend style={
    at={(0.96,0.02)},
    anchor=south east,
    draw=none,
    inner sep=1pt},
    legend style={
        font=\scriptsize,                    
        inner sep=1pt,                 
        row sep=-2pt,                  
        /tikz/every even column/.append style={column sep=2pt},
        fill opacity=0.8, text opacity=1,  
    },
    legend image post style={scale=0.6},   
    legend cell align=left
    ]
\footnotesize

\addplot[line width=0.5,color = MAT1, mark options={solid}]
table[x expr=\thisrowno{0}/1.5 ,y expr=\thisrowno{1},col sep=space,trim cells=true] {figures/surface_d3_rd6_ls6_v2.txt};
\addlegendentryexpanded{d = 3};

\addplot[line width=0.5,color = MAT2, mark options={solid}]
table[x expr=\thisrowno{0}/1.5 ,y expr=\thisrowno{1},col sep=space,trim cells=true] {figures/surface_d5_rd10_ls10_v2.txt};
\addlegendentryexpanded{d = 5};

\addplot[line width=0.5,color = MAT3, mark options={solid}]
table[x expr=\thisrowno{0}/1.5 ,y expr=\thisrowno{1},col sep=space,trim cells=true] {figures/surface_d7_rd14_ls14_v2.txt};
\addlegendentryexpanded{d = 7};

\addplot[line width=0.5,color = MAT4, mark options={solid}]
table[x expr=\thisrowno{0}/1.5 ,y expr=\thisrowno{1},col sep=space,trim cells=true] {figures/surface_d9_rd18_ls18_v2.txt};
\addlegendentryexpanded{d = 9};

\addplot[line width=0.5,color = MAT5, mark options={solid}]
table[x expr=\thisrowno{0}/1.5 ,y expr=\thisrowno{1},col sep=space,trim cells=true] {figures/surface_d11_rd22_ls22_v2.txt};
\addlegendentryexpanded{d = 11};

\addplot[line width=0.5,color = MAT6, mark options={solid}]
table[x expr=\thisrowno{0}/1.5 ,y expr=\thisrowno{1},col sep=space,trim cells=true] {figures/surface_d13_rd26_ls26_v2.txt};
\addlegendentryexpanded{d = 13};

\addplot[line width=0.5,color = MAT7, mark options={solid}]
table[x expr=\thisrowno{0}/1.5 ,y expr=\thisrowno{1},col sep=space,trim cells=true] {figures/surface_d15_rd30_ls30_v2.txt};
\addlegendentryexpanded{d = 15};

\addplot[black, dashed, line width=0.8pt]
coordinates {
    (0.103333,1e-6)
    (0.103333,1)
};
\addlegendentry{MWPM}

\end{loglogaxis}
\end{tikzpicture}%
}}
    \end{minipage}\hfill
    \begin{minipage}[t]{0.31\textwidth}
        \centering
        \subfloat[Pathfinder-BP, $d=11$, list size equal to recursive depth ($\eta=t$).\label{fig:LSeqRD}]{%
            \resizebox{\linewidth}{!}{%
                \begin{tikzpicture}
\pgfplotsset{
  set layers,
  mark layer=axis tick labels
}
\begin{loglogaxis}[
    width=\plotwidth,
    xmin=0.04,
    xmax=0.15,
    xlabel={$\epsilon$},
    ymin=1e-3,
    ymax=1,
    grid=both,
    ylabel={Logical error rate},
    legend columns=1,
    legend style={
    at={(0.96,0.02)},
    anchor=south east,
    draw=none,
    inner sep=2pt},
    legend style={
        font=\scriptsize,                    
        inner sep=1pt,                 
        row sep=-2pt,                  
        /tikz/every even column/.append style={column sep=2pt},
        fill opacity=0.8, text opacity=1,  
    },
    legend image post style={scale=0.6},   
    legend cell align=left,
    smooth,
    scaled x ticks=false,
    scaled y ticks=false,
    xtick={0.04,0.05,0.06,0.08,0.10,0.15},
    xticklabel={%
      \pgfkeys{/pgf/fpu=true}%
      \pgfmathparse{exp(\tick)}%
      \pgfmathprintnumber[fixed, precision=2, fixed zerofill]{\pgfmathresult}%
    },
    yticklabel={%
      \pgfmathparse{int(round(\tick/ln(10)))}%
      \ifnum\pgfmathresult=0 $1$\else $10^{\pgfmathresult}$\fi
    },
    ]
\footnotesize

\addplot[line width=0.5,color = MAT1, mark options={solid}]
table[x expr=\thisrowno{0}/1.5 ,y expr=\thisrowno{1},col sep=space,trim cells=true] {figures/surface_d11_rd_eq_ls_v2.txt};
\addlegendentryexpanded{$\eta$ = $t$ = 1};

\addplot[line width=0.5,color = MAT2, mark options={solid}]
table[x expr=\thisrowno{0}/1.5 ,y expr=\thisrowno{2},col sep=space,trim cells=true] {figures/surface_d11_rd_eq_ls_v2.txt};
\addlegendentryexpanded{$\eta$ = $t$ = 2};

\addplot[line width=0.5,color = MAT3, mark options={solid}]
table[x expr=\thisrowno{0}/1.5 ,y expr=\thisrowno{3},col sep=space,trim cells=true] {figures/surface_d11_rd_eq_ls_v2.txt};
\addlegendentryexpanded{$\eta$ = $t$ = 5};

\addplot[line width=0.5,color = MAT4, mark options={solid}]
table[x expr=\thisrowno{0}/1.5 ,y expr=\thisrowno{4},col sep=space,trim cells=true] {figures/surface_d11_rd_eq_ls_v2.txt};
\addlegendentryexpanded{$\eta$ = $t$ = 8};

\addplot[line width=0.5,color = MAT5, mark options={solid}]
table[x expr=\thisrowno{0}/1.5 ,y expr=\thisrowno{5},col sep=space,trim cells=true] {figures/surface_d11_rd_eq_ls_v2.txt};
\addlegendentryexpanded{$\eta$ = $t$ = 11};

\addplot[line width=0.5,color = MAT6, mark options={solid}]
table[x expr=\thisrowno{0}/1.5 ,y expr=\thisrowno{6},col sep=space,trim cells=true] {figures/surface_d11_rd_eq_ls_v2.txt};
\addlegendentryexpanded{$\eta$ = $t$ = 22};

\addplot[line width=0.5,color = black, dashed, mark options={solid}]
table[x expr=\thisrowno{0}/1.5 ,y expr=\thisrowno{1},col sep=space,trim cells=true] {figures/surface_d11_mwpm_V2.txt};
\addlegendentryexpanded{MWPM};

\end{loglogaxis}
\end{tikzpicture}%
}}
    \end{minipage}\hfill
    \begin{minipage}[t]{0.31\textwidth}
        \centering
        \subfloat[Comparison of decoding algorithms, $d=11$.\label{fig:Decoders}]{%
            \resizebox{\linewidth}{!}{%
                \begin{tikzpicture}
\pgfplotsset{
  set layers,
  mark layer=axis tick labels
}
\begin{loglogaxis}[
    width=\plotwidth,
    xmin=0.04,
    xmax=0.15,
    xlabel={$\epsilon$},
    ymin=1e-3,
    ymax=1,
    grid=both,
    scaled x ticks=false,
    scaled y ticks=false,
    xtick={0.04,0.05,0.06,0.08,0.10,0.15},
    xticklabel={%
      \pgfkeys{/pgf/fpu=true}%
      \pgfmathparse{exp(\tick)}%
      \pgfmathprintnumber[fixed, precision=2, fixed zerofill]{\pgfmathresult}%
    },
    yticklabel={%
      \pgfmathparse{int(round(\tick/ln(10)))}%
      \ifnum\pgfmathresult=0 $1$\else $10^{\pgfmathresult}$\fi
    },
    ylabel={Logical error rate},
    legend style={
        at={(0.96,0.02)},
        anchor=south east,
        font=\scriptsize,         
        draw=none, 
        inner sep=1pt,                 
        row sep=-2pt,                  
        /tikz/every even column/.append style={column sep=2pt},
        fill opacity=0.8, text opacity=1,  
    },
    legend image post style={scale=0.6},   %
    legend cell align=left,
    smooth,
    ]
\footnotesize

\addplot[line width=0.5,color = MAT1, mark options={solid}]
table[x expr=\thisrowno{0}/1.5 ,y expr=\thisrowno{1},col sep=space,trim cells=true] {figures/surface_d11_BP.txt};
\addlegendentryexpanded{BP};

\addplot[line width=0.5,color = MAT2, mark options={solid}]
table[x expr=\thisrowno{0}/1 ,y expr=\thisrowno{1},col sep=space,trim cells=true] {figures/surface_d11_BPGD_signfix_fran.txt};
\addlegendentryexpanded{BPGD};

\addplot[line width=0.5,color = MAT3, mark options={solid}]
table[x expr=\thisrowno{0}/1 ,y expr=\thisrowno{1},col sep=space,trim cells=true] {figures/surface_d11_restartBP_eta55_t5_T1-50_T2-15_fran.txt};
\addlegendentryexpanded{Restart belief};

\addplot[line width=0.5,color = MAT4, mark=square*, mark options={solid}]
table[x expr=\thisrowno{0}/1 ,y expr=\thisrowno{1},col sep=space,trim cells=true] {figures/surface_d11_RelayBP_fran.txt};
\addlegendentryexpanded{Relay BP};

\addplot[line width=0.5,color = MAT5, mark options={solid}]
table[x expr=\thisrowno{0}/1.5 ,y expr=\thisrowno{1},col sep=space,trim cells=true] {figures/surface_d11_BP_OSD_Roffe.txt};
\addlegendentryexpanded{BP OSD};

\addplot[line width=0.5,color = MAT6, mark=*, mark options={solid}]
table[x expr=\thisrowno{0}/1.5 ,y expr=\thisrowno{1},col sep=space,trim cells=true] {figures/surface_d11_rd22_ls22_v2.txt};
\addlegendentryexpanded{Pathfinder};

\addplot[line width=0.5,color = black, dashed,  mark options={solid}]
table[x expr=\thisrowno{0}/1.5 ,y expr=\thisrowno{1},col sep=space,trim cells=true] {figures/surface_d11_mwpm_v2.txt};
\addlegendentryexpanded{MWPM};

\end{loglogaxis}
\end{tikzpicture}%
}}
    \end{minipage}

    \vspace*{-2.5mm}
    \caption{Logical error rate versus error probability $\epsilon$ for planar surface codes.}
    \label{fig:LER_overview}
\end{figure*}
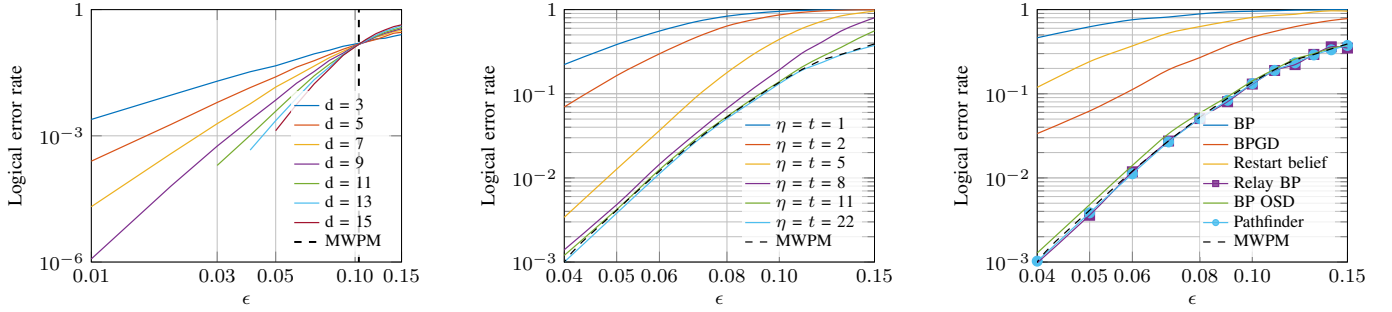

We present numerical results for Monte Carlo simulations of the logical error rate for planar surface codes of different sizes under independent phase-flip errors, see Section~\ref{sec:err_model}. 
Accordingly, only $\bm{H}_X$ is used in the simulations. By the $X$/$Z$ symmetry of the CSS structure, the results apply equally to bit-flip errors. 
A decoding attempt is counted as successful if the residual error $\bm{e} \oplus \hat{\bm{e}}$ is a stabilizer. 
Shots for which no candidate is found are counted as logical errors. Each Monte Carlo point is obtained by collecting $100$ logical errors. For Pathfinder-BP we use $T_1=50$ and $T_2=15$ throughout, with a flooded (fully parallel) BP schedule. 

Figure~\ref{fig:threshold} shows the \ac{LER} as a function of physical error rate for the Pathfinder-BP with a list size and recursive depth both set equal to two times the code distance, $2d$. 
As can be observed, Pathfinder-BP achieves a threshold\footnote{The threshold is estimated as the crossover of the logical error rate curves for different distances. The value $\sim 0.103$ is consistent with the finite-size crossover for planar surface codes with open boundaries.} comparable to \ac{MWPM}, $\sim 0.103$. 

Figure~\ref{fig:LSeqRD} shows the \ac{LER} for the $d=11$ surface code under
\ac{MWPM}, which serves as a benchmark, and under Pathfinder-BP with different
parameters. Due to space limitations, we set the list size equal to the
recursion depth, $\eta = t$, and vary it from 1 to $2d = 22$.
As expected, the performance improves with increasing $t$ and $\eta$. For
$\eta = t = d = 11$, Pathfinder-BP performs on par with \ac{MWPM}.

Figure~\ref{fig:Decoders} compares the \ac{LER} versus the physical error rate
for: (i) \ac{BP} with $50$ iterations; (ii) \ac{BP} with guided decimation,
$T_{\text{BP}} = 25$; (iii) restart belief\footnote{We used QBelief, the
implementation in~\cite{valentini2025}.} with $T_1 = 50$, $T_2 = 15$,
$\eta = 55$ and $t = 5$; (iv) relay \ac{BP}\footnote{We used IBM's
implementation~\cite{muller2025improved}.} with $R=601$ ensemble elements, $T_1=80$, 
$T_2 = 60$, $S = 5$ solutions sought, and memory weights drawn uniformly from
$(-0.24, 0.66)$; (v) \ac{BP}-\ac{OSD} of order $\ell = 2$; (vi) Pathfinder-BP
with $T_1 = 50$, $T_2 = 15$ and $\eta = t = 22$; and (vii) \ac{MWPM}.
Pathfinder-BP and relay \ac{BP} outperform all other \ac{BP}-based decoders
and perform on par with \ac{MWPM}.

In terms of \ac{LER}, Pathfinder-BP and relay \ac{BP} are on par; their
latencies, however, differ substantially.
In relay \ac{BP}, each ensemble element is initialized with the posteriors of
the previous one, so the elements run sequentially and the number of \ac{BP}
iterations is bounded by $T_1 + R\,T_2 = 36140$, whereas the parallel structure
of Pathfinder-BP bounds it by $T_1 + t\,T_2 = 380$.
Table~\ref{tab:latency} reports the empirical distributions at $\epsilon = 0.04$, for the same parameter sets used for Figure~\ref{fig:Decoders}.
On average, Pathfinder-BP requires $102$ iterations and relay \ac{BP} $120$.
At the $90$th percentile, relay \ac{BP} is faster ($195$ vs.\ $326$),
since its first ensemble element mostly converges.
However, its distribution is heavy-tailed: the $99$th and $99.9$th
percentiles are $830$ and $6528$, and  $380$ iterations are exceeded in $2.87\%$
of the instances.
Moreover, the choice $\eta = t = 22$ is conservative for Pathfinder-BP. As shown in
Fig.~\ref{fig:LSeqRD}, lower values of $t$ also allow to perform on par with
\ac{MWPM}, which lowers the worst-case bound.
Since a real-time decoder must keep pace with syndrome extraction, the worst
case, not the average, determines the admissible latency. This makes
Pathfinder-BP the better choice for real-time decoding.

\begin{table}[t]
  \centering
  \caption{Number of \ac{BP} iterations per decoding instance at $\epsilon = 0.04$.
    Empirical statistics over $10^{5}$ instances.}
  \label{tab:latency}
  \begin{tabular}{lcc}
    \toprule
                                   & Relay \ac{BP} & Pathfinder-BP \\
    \midrule
    Mean                           & $120$   & $102$   \\
    $90$th percentile              & $195$   & $326$   \\
    $99$th percentile              & $830$   & $380$   \\
    $99.9$th percentile            & $6528$  & $380$   \\
    Worst case                     & $36140$ & $380$ \\
    $\Pr[\text{iterations} > 380]$ & $2.87\%$ & $0$  \\
    \bottomrule
  \end{tabular}
\end{table}

\section{Conclusions and Discussion}
We have proposed Pathfinder-BP, a recursive decoder for topological
quantum codes that augments belief propagation with a controlled
bit-flipping strategy. 
Under code-capacity noise, Pathfinder-BP achieves a noise threshold and
logical error rates on par with \ac{MWPM} on planar surface codes.
A key advantage of Pathfinder-BP is that it does not require performing a
matrix inversion, which is complex and unsuitable for hardware
implementation. 
For surface codes, and assuming parallel hardware, the worst-case latency of Pathfinder-BP is quasi-linear in the code distance, whereas the worst-case latency of matching decoders is superlinear.
This makes Pathfinder-BP a strong candidate for low-latency hardware
implementations.

Pathfinder-BP applies to arbitrary CSS stabilizer codes and not just to surface codes. In future work, we will extend Pathfinder-BP to other code families, including color, hypergraph product codes and BB codes and to more realistic noise models, such as correlated and circuit-level noise.

\bibliographystyle{IEEEtran}
\bibliography{IEEEabrv,references}

\flushcolsend
\end{document}